\documentclass{iaus}
\usepackage{graphicx}

\title[JD 11.~~The Thick Disk] 
{The Galactic Thick Disk: An Observational Perspective}

\author[Bacham E. Reddy]   
{Bacham E. Reddy}

\affiliation{Indian Institute of Astrophysics, Bengaluru, 560034,
\\ email: {\tt ereddy@iiap.res.in}} 

\pubyear{2009}
\volume{265}  
\pagerange{119--126}
\jname{Chemical Abundances in the Universe: Connecting First Stars to Planets}
\editors{K. Cunha, M. Spite \& B. Barbuy, eds.}
\begin{document}

\maketitle

\begin{abstract}
In this review, we present a brief
description of observational efforts to understand the Galactic thick disk and its relation
to the other Galactic components. This review primarily focused on 
elemental abundance patterns of the thick disk population to pin down the process or processes that
were responsible for its existence and evolution.
Kinematic and chemical properties of disk stars establish that the thick disk
is a distinct component in the Milky Way. The chemical enrichment and star formation 
histories hold clues to the bigger picture of understanding the Galaxy formation.
 
\end{abstract}

\firstsection 
\section{Introduction}

Deciphering the history of the birth and the growth of the Milky Way Galaxy is one of the major
outstanding astrophysical problems.
Detailed studies
of our Galaxy would help to gain insights into the formation 
and evolution of other
galaxies, and thereby large scale structure formation
in the universe.    
As inhabitants,
we have much better access to our Galaxy to resolve  its building blocks: the stars.
The properties of stars that constitute the Galaxy are clues to the processes  
involved
in the making of the Milky Way Galaxy we see today. The observational data on stellar motions and
 photospheric abundances played decisive roles in the progression of our understanding of the
Galaxy. The major, perhaps the first, quantitative study of stellar kinematic and 
chemical properties of a large sample of nearby dwarfs
by Eggen, Lynden-Bell and Sandage (1962), known as ELS, laid the foundation for the Galaxy
formation models.  In this seminal paper ELS showed that the
eccentricity of stellar orbits and the photometrically derived UV excess ($\delta$ (U-B)), an indicator
of metallicity, are correlated: stars with the largest excess (metal-poor)  are 
found to be moving
in highly elliptical orbits, where as the stars with little or no excess (solar metallicity stars)
are found to have circular orbits. ELS attributed this to the rapid collapse of a proto-galactic cloud
forming the halo quickly, and the rotationally supported disk component thereafter. Further studies,
based on improved observations of globular clusters in the Galaxy, led to the fact that mergers play
significant role in the evolution of the Galaxy \cite[(Searle $\&$ Zinn 1978)]{SearleZinn1978}. Thanks to large scale
surveys of both the kinematic and photometric properties of stars, we now know that the Galaxy consists
of four major components: the halo, the thick disk, the thin disk and the bulge.   
There are excellent reviews
which trace the progressive understanding of the subject 
\cite[(e.g; Gilmore 1989; Majewski 1993; 
Freeman $\&$ Bland-Hawthorn2002)]{majewski93}. In this review, we primarily concentrate on the thick disk component
and its relation to the Milky Way galaxy.

\section{The Galactic Thick Disk}

Evidence for a distinct component,the so called thick disk, in the disks first
came from the data of external galaxies.
Studies of brightness distribution as a function of
distance from the galactic planes of some of the edge-on galaxies (NGC 4570, NGC 4350) 
required fitting of a third component, in addition to the
usual two components (a thin disk and a bulge), for the fainter
part (Burstein 1979), termed as the "thick disk" for its diffused 
or fluffy appearance. Shortly, thereafter
a definitive evidence of the existence of the thick disk in the Milky Way Galaxy had emerged
from the photometric data of a large sample of stars towards the South 
Galactic Pole (Gilmore $\&$ Reid 1983). From the data of star counts 
or stellar density distribution
as a function of  distance from the Galactic plane, two distinct components were identified:
one with a scale height of 300 pc within 1 kpc of the plane, and the second 
with a scale height of 1350 pc which dominates the stellar density law in the Galaxy between
1 kpc to about 5 kpc. The first one with a smaller scale height was identified as 
the old disk or the thin disk and  
the second component with a large scale height was identified as the thick disk 
analogous to the one found in the
edge-on galaxies. 
The thick disk population is the same population of stars in the Galaxy 
that was
known earlier, but not so well defined, as the intermediate population 
II or IPII (Str\"{o}mgren 1964). Subsequent studies of photometric and metallicity data
revealed that the thick disk population stars are metal-poor, old and show no evidence of 
metallicity gradients within the thick disk (e.g; Gilmore, Wyse $\&$ Jones 1995). 
Thus, the concept of the thick disk is established and is 
termed as a major modification
to the ELS model of rapid collapse for the formation of the Galaxy. 

\section{Global properties}

Identification of Galactic stellar populations into distinct components 
relies, primarily, on five basic differences: location, structural properties, kinematic
motions, age and chemical make-up.
Until late 90s,  studies mostly
focused on defining the global properties of the thick disk like scale height, scale length, local
normalization, metallicity distribution, kinematic properties and to some extent ages 
relative to the other components.
Thanks to the Hipparcos space mission for accurate stellar astrometry, and the major ground based
surveys like Radial Velocity Experiment (RAVE), Sloan Digital
Sky Survey (SDSS), and 2 Micron Sky Survey (2MASS) it became abundantly clear that the
thick disk exists and it has distinct global properties which are different
from that of the thin disk, halo and the bulge. Since the halo and the bulge 
populations are known to  differ both in morphology 
and kinematics from the disk populations, we concentrate on distinguishing
two distinct populations within the Galactic disk: the thin disk and the thick disk.

Recently measured values of the thick disk population stars are
summarized in Table~1. The thick disk has scale height in the range of 500pc to 1100 pc
which is two to four times of the thin disk (300 pc),and
its metallicity peaks at [Fe/H] $\approx$ $-$0.6 dex compared to the thin disk which
peaks at [Fe/H]$\approx$ $-$0.2. Thick disk stars are older (8-12 Gyrs) compared to their
counter parts in the thin disk stars ($<$10 Gyrs). Local normalization of the
thick disk population relative to the thin disk population varies from survey to
survey and which is in the range of 2$\%$  to 13$\%$. That is there are 2 to 13 thick disk stars
for every 100 thin disk stars in the solar neighborhood. Volume limited  (stars
within the radius of 25 pc from Sun) survey by Fuhrman (2008) suggests even much
larger proportion (20$\%$) of thick disk stars in the local neighborhood implying
a massive thick disk. The population of the thick disk stars increases 
as one moves away from the Galactic plane in the z-direction. The thick disk
dominates other populations
beyond ~1.0 kpc. If the thick disk is a separate component
one would also expect distinct 
kinematic motion ($U$, $V$, $W$) for the thick disk population which is different from the
other two components: the halo and the thin disk.

One of the potential discriminators of the Galactic components is their
average rotational velocity (V$_{rot}$) or asymmetric 
drift velocity ($V_{lag}$) with respect to the 
Galactic rotation of 220 km s$^{-1}$.
This has been known for a while (see Majewski 1993). 
As shown in Table~1, values of $V_{lag}$ for the thick disk range anywhere between
20 km s$^{-1}$ (Chiba $\&$ Beers 2000) to 
51 km s$^{-1}$ (Soubiran et al. 2003). Large range in the measured $V_{lag}$
is, perhaps, due to the sample selection in the respective studies. However,
$V_{lag}$ of 40 - 50 km s$^{-1}$ is generally in good agreement with  
the other recent studies (e.g; Robin et al. 2003).
Compare this with the halo population which has  V$_{lag}$ $\approx$ 220 km s$^{-1}$  (or statistically
stationary w.r.t to the Galactic rotation) and with the thin disk of V$_{lag}$ $\approx$
5-10 km s$^{-1}$ (Robin et al. 2003). The measured velocity 
ellipsoid ($\sigma_u$, $\sigma_v$, $\sigma_w$)  of the thick disk is about twice that
of the thin disk and half of the halo values. In all of the measurements, Gaussian distribution
is assumed, and the non-Gaussian velocity distributions can't be ruled out.  

In summary, as far as the Global properties of the thick disk are concerned, 
measurements based on different data sets and different methodologies
agree with each other suggesting a level of overall confidence in characterizing the
thick disk. Discontinuity in the luminosity function, metallicity distribution
and the difference in rotational velocity are some key evidences
for the distinct nature of the
thick disk population in the disk of the Galaxy.

\begin{table}
  \begin{center}
  \caption{Summary of the Global and Kinematic properties of the Thick Disk}
  \label{tab1}
 {\scriptsize
  \begin{tabular}{lcccccc}\hline 
{\bf Parameter} & {Carollo et al.} & {Juric et al.} & {Siegel et al.} \\
                & {(2009)} & {(2008})  & (2002)   \\
\hline
Scale height(kpc) &0.51$\pm$0.04    & 900 pc    & 940 pc  &          \\
Scale length (kpc) & 2.20$\pm$0.35    & 3.6 kpc   & ...  &          \\
Local normalization & 15$\pm$7$\%$   & ...   & 8.5$\%$      \\ \hline 

                    & {Carollo et al.} & {Soubiran et al.} & {Chiba $\&$ Beers}  \\ \hline
 V$_{lag}(km s^{-1}$ & 38$\pm$2    & 51$\pm$5   & 20   &   &   &    \\
 Velocity Ellipsoid & (53$\pm$2,51$\pm$1, 35$\pm$1) & (63$\pm$6,39$\pm$4, 39$\pm$4)   
& (46$\pm$4,50$\pm$4, 35$\pm$3)  &   &       \\
  Metallicity peak & $-$0.6   & $-$0.48   &   &   &       \\
\hline
  \end{tabular}
  }
 \end{center}
\vspace{1mm}
\end{table}

\section{Stellar Abundances of the Thick Disk} 

Low mass main sequence dwarfs play an important role in tracing the 
sequence of events that took place during the Galaxy evolution. Unevolved
dwarfs preserve the chemical composition of their natal clouds. Thus, the determination of
photospheric abundances and their relative ratios is of great importance, in other
words, chemical tagging of individual stars would help to understand the fundamental
processes involved in the making of the Galaxy. 
In their review,  Freeman $\&$ Bland-Hawthorn(2002)
stressed the importance of the differential abundance analysis for accurate abundance
determinations. 
Sample selection
is one of the important steps in this endeavour to keep the uncertainties at minimum.
Dwarfs of spectral type F and G are well suited for
accurate
abundance determinations. 
In the stars of earlier spectral type, spectral lines are 
very broad due to higher rotational velocity, and  some of them
are either very weak or absent due to hotter temperatures. On the other hand, cooler stars
have rich spectra with crowded regions making it difficult to analyze, 
and hence larger uncertainties.
Another important consideration is the selection of elements. The 
$\alpha$-elements like O, Mg, Si, Ca or Ti are thought to be predominantly produced
in the short lived massive star explosions (SNII) 
  and elements like Fe, Ni etc. are the main products from the long lived low mass star
explosions (SNIa).  SNII are more frequent in the early epochs and SNIa starts contributing
much later. Thus, 
ratio of [$\alpha$/Fe] is a tracer of the
chemical evolution and star formation rates in the Galaxy as well as a indicator of relative frequency of
SNII to SNIa explosions.

The abundance study of the Galactic disk stars of Edvardsson et al. (1993) 
is the first major work in this direction. They have analyzed 
high resolution spectra of 189 F- and G- dwarfs and measured 
abundances of 12 elements, which provided important constraints for the 
Galactic chemical evolutionary models. 
They have noticed significant scatter 
in [$\alpha$/Fe] versus [Fe/H], particularly, below [Fe/H] $\approx$ -0.4. 
Stars with relatively higher [$\alpha$/Fe] ratios found to be preferentially more eccentric and closer
to the Galactic center. This was attributed to an efficient star formation 
rate in the inner disk. Surprisingly, the possibility of the thick disk population to 
the observed scatter in the abundances was
never raised in the entire lengthy paper. By the late 1990s, it became increasingly clear
that the thick disk harbours a stellar population whose chemical history is distinct from the
rest of the components in the Galaxy, particularly the thin disk. Fuhrman (1998) study of carefully
selected and unbiased, but volume limited, sample of F and G dwarfs indicated that stars with
intermediate kinematic properties show higher [Mg/Fe] ratio compared to the thin disk 
stars in the overlapping metallicities. This was confirmed from the detailed study of many more elements
but for a small kinematically pre-selected sample of 10 stars (Prochaska et al. 2000). 
Abundance ratios of 
[$\alpha$/Fe]
found to be
higher compared to the thin disk stars but very similar to halo and bulge stars. 
Contrary to the above results,
Chen et al. (2000) study of pre-selected
samples of thick disk showed continuous chemical evolution from the thick to
the thin disk populations.

\begin{table}
  \begin{center}
  \caption{Summary of a few key abundances ratios for the kinematically selected
stars the Halo, thick disk, MWTD, and
 thin disk. Data is adopted from a series of papers by Reddy et al. group.}
  \label{tab1}
 {\scriptsize
  \begin{tabular}{lcccc}\hline 
{\bf [X/Fe]} & Halo & MWTD & Thick Disk & Thin disk \\
\hline
{[Mg/Fe]} &  0.31$\pm$0.06  & 0.32$\pm$0.07 & 0.31$\pm$0.09  & 0.09$\pm$0.05  \\
{[Si/Fe]} & 0.23$\pm$0.06   &  0.27$\pm$0.09 & 0.23$\pm$0.15 & 0.07$\pm$0.04\\
{[Ni/Fe]} & 0.02$\pm$0.04   & $-$0.08$\pm$0.09 & $-$0.08$\pm$0.11 & $-$0.02$\pm$0.02 \\
{[Eu/Fe]} & 0.38$\pm$0.13   &  0.44$\pm$ 0.24 & 0.35$\pm$ 0.15 & 0.12$\pm$0.10 \\
\hline
  \end{tabular}
  }
 \end{center}
\vspace{1mm}
\end{table}

\section {Evidence for the distinct chemical history}

Is the thick disk a discrete entity in the Galaxy? Does it evolve from the thin disk
to the thick disk or vice versa? What is the metallicity range of the thick disk? Answers
to these questions will help to piece together the puzzle of the thick disk and its relation
to the rest of the Galaxy. As noted in the above section, 
there seems to be evidence that the chemical history of the thick disk is 
different,
at least from that of the thin disk. But this needs to be validated 
from the larger samples
of stars and more number of elements of different nucleosynthetic history. Also,
to put the thick disk in the context of the Galaxy evolution, uniform study of samples
of stars from other components is essential. Precisely, this was pursued 
independently, in the 
recent studies (Reddy et al. 2003, Reddy, Lambert, Allende Prieto 2006,
Reddy $\&$ Lambert 2008; 
Bensby et al. 2003,2005,2007; Fuhrman 1998, 2008).
Bulk of the information on the 
abundances of the thick disk and thin disk populations is drawn from  
the above studies particularly from the two groups: Reddy et al and Bensby et al.
These studies  are complimentary  to each other  as the sample stars of Bensby et al. are
drawn from the southern hemispheres whereas  the samples in the studies of Reddy et al. 
are from the northern
hemisphere. Common stars among the two studies are very few. However, the agreement between the
two results,in most cases, is extremely good indicating the maturity of the different 
model atmospheres and the analysis techniques. 

\begin{figure}[b]
\begin{center}
 \includegraphics[width=4.4in]{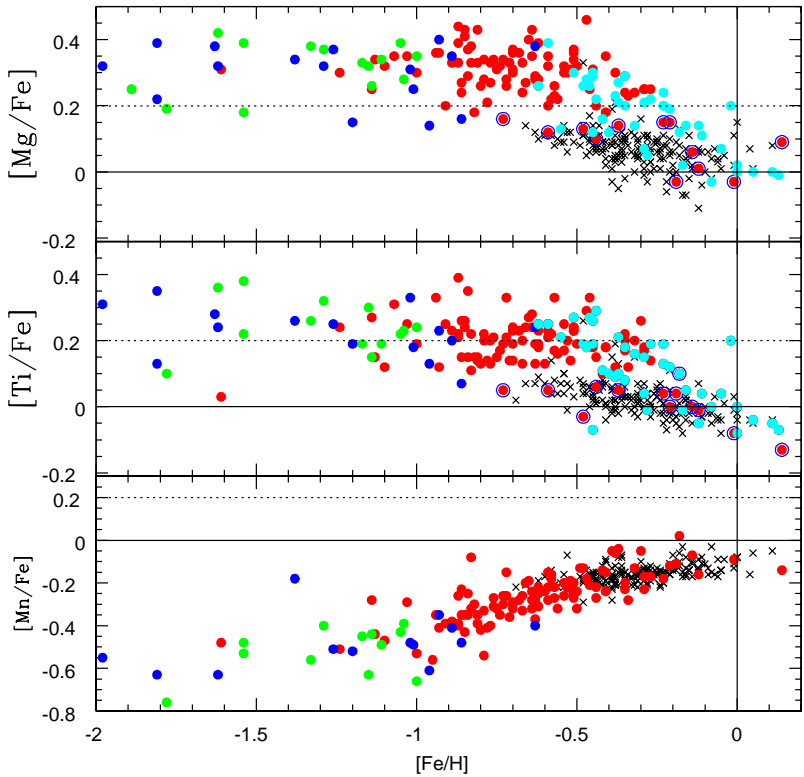} 
 \caption{Abundances ratios [Mg/Fe], [Ti/Fe] and [Mn/Fe] versus [Fe/H] for samples
of the thick disk (red circles), the thin disk (black crosses) 
and the halo stars (blue circles). Light blue and green circles are shown for stars 
of metal-weak and metal-rich thick disks, respetively. TKTA stars are shown as red circles embedded
in blue circles.}
   \label{fig1}
\end{center}
\end{figure}

In the studies of Reddy et al. (2003, 2006) a carefully pre-selected sample,
based on kinematic properties, of
about 400 nearby dwarfs belong to the thin disk, the thick disk, or the halo 
were subjected to high resolution spectroscopy. Stars were grouped into the thin disk, thick disk
and the halo populations based on their kinematic definitions and relative fraction
of each of these populations in the solar neighbourhood.  
Quantitative abundances of 22 to 26 elements were determined with  uncertainties 
as low as 0.04 - 0.06 dex for most of the key elements. The mean [X/Fe] ratios are given
in Table~2 for the thick disk, thin disk and the halo stars. 
Results for a few key elements
are shown in Figure~1.  Similar studies based on similar number of stars were carried out 
in a series of papers by Bensby et al. group. In general, results from both the studies are
in good agreement.
Some of the Key conclusions from these results are: a)
Abundance ratios of elements that are known to be the dominant yields from SNII
are clearly larger than
their counter parts in the thin disk in the overlapping metallicity 
($-$0.3 - $-$0.8 dex) implying a distinct chemical history for the thick disk from that of the thin disk.
However, the mean abundance ratios of elements of thick disk stars are
indistinguishable from those of MWTD or halo stars and even the bulge stars ( e.g; Mel\'{e}ndez et al. 2008). 
The similarity of the thick disk abundance ratios with that of the halo and bulge may suggest 
that the three components are similarly old and had a rapid star formation.
b) The run of [$\alpha$/Fe] with [Fe/H] for stars of the thick disk
shows little or no slope
against [Fe/H] suggesting rapid formation of the thick disk. Slight slope towards higher
metallicities is attributed to $\alpha$-element yield
dependence on the initial metallicities of SNII. 
 c) Another
important outcome of the analysis is that the star-to-star scatter 
in the trends which is comparable to the scatter expected from the measurements
indicating stars both in the thin as well as in the thick disk formed from the well mixed
gas, and  d) Age estimates indicate that the stars of the thick disk are 
older on average (8-12 Gyrs)
compared to the thin disk stars (t $<$ 10 Gyrs).  Age results show that 2-3 Gyrs elapsed between the
first stars of the thin disk and the first stars of the thick disk suggesting sufficient time for
the SNIa role in the chemical history of the thick disk.

\section{Metal-Rich end of the thick disk}
Is thick disk evolving similar to the thin disk or is it a frozen entity in the
Galactic disk? Evidence of delayed SNIa contribution in the thick disk was first proposed
by Bensby et al. (2003).
The claim of the so called "Knee" connecting thick disk stars 
from high [$\alpha$/Fe] trend
below [Fe/H] = $-$0.3 to thick disk stars of solar [$\alpha$/Fe] 
values above [Fe/H] = $-$0.2 was disputed first by Reddy et al. (2006) and 
later by Ram\'{i}rez et al. (2007). Confirmation of presence or absence of the "knee" 
in the [$\alpha$/Fe] trend with [Fe/H], and its smooth merging with the thin disk trend
is of significance. Such an evidence will constrain the thick disk 
chemical enrichment models.

Evidence of SNIa contribution to the chemical
history of the thick disk was not very clear in the Reddy et al. (2006) abundance survey
of the thick disk. This was partly due to the 
presence of a different population
with thick disk kinematics but with thin disk chemistry (Reddy et al. 2006) and partly 
due to lack of sufficient number of stars to form
the "knee". As shown in Fig 1 (red symbols: Reddy et al. 2006), one could see a hint
of drop in the [$\alpha$-elements/Fe] trends at [Fe/H] = $-$0.35 and a few thick disk stars 
above [Fe/H] $\approx$ $-$0.2 with very similar abundances of thin disk. Evidence for the
drop in the [$\alpha$/Fe] was overlooked to explain the thick disk stars 
below 
[Fe/H] = $-$0.35 with thin disk abundances. Together they were termed as TKTA stars
(Fig 1: red symbols embedded in
blue circles) 
for thick disk kinematics and thin disk abundances. Ram\'{i}rez et al. (2007) explored the
thick disk chemical evolution using O abundance of about 500 dwarfs. 
From the results of detailed LTE and NLTE analysis they have suggested 
that the thick disk chemical evolution
ends at [Fe/H] $\approx$ $-$0.3 dex, and the proposed existence of a 
knee in the O abundance trend
was challenged. This required further studies to settle the issue.

Bensby et al (2007) undertook a systematic study of thick disk stars in the metallicity
range of $-$1 $\leq$ [Fe/H] $\leq$ $+$0.4 with sufficient number of thick disk stars
above [Fe/H] = $-$0.35. Results showed strong evidence that the thick disk extends
at least to the solar metallicities and the drop in 
the [O/Fe] pattern occurred at [Fe/H = $-$0.35. This was also confirmed from the new
sample based on several elements (Reddy et al. 2009 and Bensby et al. in this volume).  
In Figure~1, our new kinematically chosen thick disk sample stars 
are shown in light blue. They connect 
thick disk stars (red circles) of high [$\alpha$/Fe] trend with thin disk stars of solar
[$\alpha$/Fe] trend.  
Results in Fig~1 and Fig~2 highlight the complexity of the disk structure:
the TKTA population persists in the new
analysis, confirming the earlier claim by Reddy et al. (2006). 
The results in Fig~2 also suggest that there seems to be multiple
channels that are connecting thick disk trends with that of 
the thin disk.

\section{Metal-Weak end of the thick disk}
There are not many studies that focused on the Metal Weak Thick Disk stars (MWTD). Morrison et al. (1990) were the first to point
out stars that are very similar to the disk kinematics  
but significantly metal-poor, [Fe/H] $<$ $-$1.0. Later, this turned out to be not real. The
recalibration of the metallicities suggested that the so called metal weak thick 
disk stars  
were in fact found to have
metallicities [Fe/H] $>$ $-$1.0 (Ryan $\&$ Lambert 1995; Twarog $\&$ Anthony-Twarog 1996). 
Stronger evidence for the MWTD
appeared in the analysis of large data sets of metal-poor stars
(Chiba $\&$ Beers 2000). The [M/H] versus rotational velocity relations suggested stars as metal-poor
as [M/H] = $-$1.7 have disk like velocities of $V_{lag}$ $>$ 70 km s$^{-1}$ slightly higher
than the values estimated for the canonical thick disk and much lower 
than the halo population ($V_{lag}$ $\approx$ 220 km s$^{-1}$) (see Table 1). 

Isolation of the MWTD and determination of its status either as a metal-poor tail of the
thick disk, or a collection of halo stars or a discrete Galactic component is important
and worth to pursue to gain insights into the early epochs of the disk formation.  Establishing
the MWTD thick disk as a discrete entity would provide  another evidence for the hierarchical
galaxy formation. On the other hand, establishing it 
as a metal-poor tail of the thick disk would tell us the 
star formation history of the thick disk. 
An attempt was made to systematically search for MWTD and study their 
chemical abundance pattern in relation
to the thin disk, thick disk and halo populations (Reddy $\&$ Lambert 2008). For this study they
made use of the two catalogues (Arifyanto et al. 2005; Schuster et al. 2006) to search
for MWTD. From the combined catalogue of about 1700 stars,  they have 
chosen the criteria for the MWTD: a) [M/H] $<$ $-$1.0, (b)  
rotational velocity V$_{r}$ $>$ 100 km s$^{-1}$ (or $V_{lag}$ $>$ 120 km s$^{-1}$), (c) 
$|W_{LSR}|$ $\leq$ 100 km s$^{-1}$ and $|U_{LSR}|$ $\leq$ 140 km s$^{-1}$. First condition
eliminates contamination of stars from the thin disk and conditions b and c significantly
enhance the proportion of MWTD to the halo stars. 

High resolution abundance analysis study showed
no clear difference in any elemental abundance trends, among several elements
examined, between stars of the MWTD, halo or the canonical thick disk (see Table 2). 
This raises three possibilities; a) MWTD is a metal-poor end of the thick disk 
and shows no distinguishable chemical abundance
difference as the thick disk formed more rapidly, b) MWTD is a discrete entity with indistinguishable
abundances either from the thick disk or halo populations. 
The difference in age between the three populations is too small ( 1-2 Gyrs) to reflect differences
in their chemical composition, and c) finally MWTD stars are basically a collection of 
halo stars with disk like kinematics. None of
these possibilities could be ruled out at present as far as the MWTD is concerned. To establish the MWTD
status, one may require large data sets with reliable kinematics from accurately measured distances,
proper motions and radial velocities. Some of these efforts are underway. In a recent study Carollo et
al. (2009) explored the full space motions 
for a sample of about 17000 stars in a local volume within 4 kpc of the Sun. Photometric, low resolution
spectroscopic, and
proper motion data were taken from SDSS. In the metallicity 
regime of $-1.7 < [Fe/H] < -1.0 $ and in the interval of vertical distance from the Galactic
plane $1 < |Z| < 2 $ kpc, observational data required to account for two distinct populations: inner
halo ( V$_{rot}$ $\approx$ 0) and MWTD with 
V$_{rot}$ $\approx$ 100 - 150 km s$^{-1}$ (or V$_{lag}$ $\approx$ 120 - 70 km s$^{-1}$).
They suggest that the MWTD is a separate component and is different from the canonical thick disk.
This is a step forward to resolve the issue. However,  this is based on a single paramete i.e.,
rotational velocity lag between the two thick disk stars. The presence of a large kinematic and spacial
overlapping with the halo and the canonical thick disk, and the expected large uncertainties in 
derived astrometry will require additional studies to fully resolve the status of MWTD.
Accurate astrometry for large samples and followed by accurate abundance study would
certainly help to unravel the origin thick disk over its full metallicity range.

\begin{figure}[b]
\begin{center}
\includegraphics[width=4.4in]{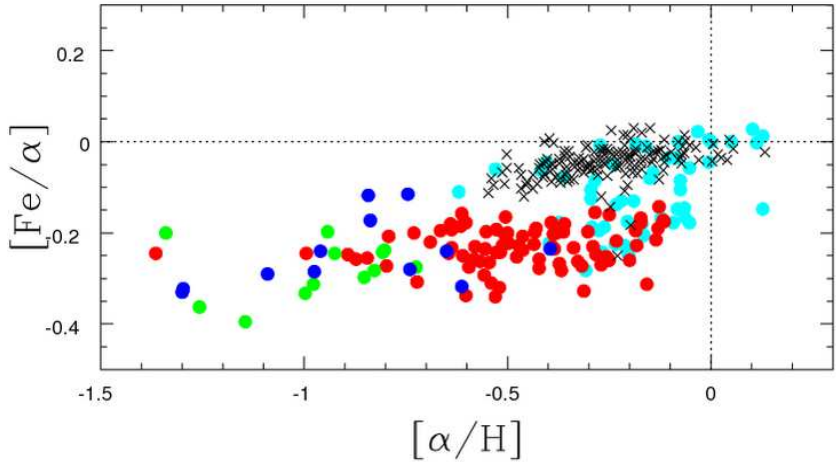} 
 \caption{Abundances ratio [Fe/$\alpha$] versus [$\alpha$/H] for samples
of the thick disk, the thin disk and the halo stars. Colour and symbol coding are same as 
those for in the Figure~1.}
\label{fig1}
\end{center}
\end{figure}

\section{Thick disk in the context of the Galaxy formation}
In the last two decades, a wealth of observational information was collected
not only on the thick disk of the Milky Way Galaxy but also on the thick disks of
many edge-on galaxies (e.g; Yochin $\&$ Dalcanton 2008). 
Observational
results deduced from various data sets by different groups are consistent, and
they put stringent constraints on the
theoretical modeling of the thick disk origin. To put the thick disk in the 
overall perspective
of the Milky Way Galaxy formation,
theoretical modelling has to take into
account all the well established observational parameters: a) scale heights,
b) local relative density of the thick disk,
d) significantly older (8-12 Gyrs) age of thick disk stars
compared to the thin disk
stars of similar metallicity, d) distinct kinematic properties, e) no noticeable
velocity as well as metallicity gradients as a function of vertical distance or age, f) distinct
metallicity distribution: thick disk peaks at $-$0.6 dex, thin disk at $-$0.2 dex and the halo at
$-$1.6 dex, g) distinctly higher [$\alpha$/Fe] and [Eu/Fe] ratios from that of the thin disk, 
h) little or no 
gradient in the [$\alpha$/Fe] over a large range of metallicity, $-$1.2 $<$ [Fe/H] $<$ $-$0.3, i)
drop in
the [$\alpha$/Fe] ratios at [Fe/H] $\approx$ $-$0.35 dex and smooth merging with the thin disk
ratios at about solar metallicities, j) very little/no cosmic scatter in the abundance ratios
over the large [Fe/H] range, k) existence of kinematically thick disk stars with thin disk abundance
trends over the full metallicity range of the thin disk.  Unfortunately, the status of the metal-poor
end of the thick disk is yet to be understood,otherwise, it would have been 
an important constraint.

Ever since the existence of thick disk was established in disk galaxies as well as in the Milky Way,
variety of theoretical models were proposed to explain the process that was responsible for the
thick disk.
Broadly, one could group these into two categories: top down and bottom up models.
Briefly, in the top down models, thick disk is formed either through a rapid or dissipative collapse
of protogalactic clouds. In this scenario halo forms first, then the thick disk, and then the
thin disk. Top down models, by their nature, predict gradients both in abundances and kinematics, and
do not predict distinct chemical abundances of the thick disk
from the thin disk over the overlapping metallicities. Top down models fail to meet many of
the well defined observational parameters listed above. 
For an exhaustive list of different models and their full description, 
readers are advised to refer to Majewski (1993) and references therein. 

In the case of  
bottom up models thin disk already exists, and the thick disk forms later either 
through satellite mergers, accretion of debris or through radial mixing. These models
are successful in predicting many of the observed properties of the thick disk. It may
not be exaggerating to say that all the three possibilities may have contributed     
to the formation of the thick disk. Simulations of galaxy formation through hierarchical
build up based on the
$\Lambda$CDM universe are regularly reported in the literature. Many of these simulations
predict emergence of thick disk in the galactic disk, and some of its distinct properties 
from the thin disk and the halo. Quinn, Hernquist $\&$ Fullagar (1993) proposed that 
satellite mergers heat the preexisting thin disk but not the gas.
Thick disk stars are primarily the fossils of the thin disk. These models predict 
some of the distinct global properties like scale height and kinematic properties of thick disk
stars. Later thin disk forms from the reformed gas from the merged satellites.
Models predict only mild reduction in the rotational velocity, and do not account for
the chemical enrichment of the thick disk from the delayed SNIa. Thin disk heating by
mergers was probed further ( Kazantzidis et al. 2008)  
by including massive satellites  
to heat the thin disk sufficiently to form the thick disk. Massive 
satellites were more frequent at redshifts above z$\approx$1 which corresponds 
to the age of the thick disk in the
$\Lambda$CDM.
Alternatively, Abadi et al. (2003)  simulations of Galaxy formation in $\Lambda$CDM universe,
resulted a thick disk consisting mostly of stars from the satellite debris and a few through accretion. 
A clear distinction between dynamically cold disk of stars on nearly circular orbits and a thicker
disk with orbital parameters in between the thin disk and the halo emerged. Results help to explain
the existence of non-negligible fraction of thin disk stars which pre-date the last 
major merger history. However, the chemical evolution of the thick disk in the simulations
in this framework need to be probed.

In a series of papers Brook et al. (2007 and references therein) proposed the formation of
thick disk through heating of the thin disk by gas rich mergers and starburst during 
the merger process. Starburst during the merger accounted the observed thick disk properties
like kinematic dispersions, higher [$\alpha$/Fe], and rapid star formation. Rapid star
formation ensures higher [$\alpha$/Fe] even at higher metallicities with significantly
much older age compared to thin disk stars with similar metallicities but much younger.
The models also predict
very low velocities for the merged stars and some of them have counter rotation. 
Though, simulations predict trend of decreasing of [$\alpha$/Fe] with [Fe/H] but stops at much
higher [$\alpha$/Fe]$\approx$ 0.2 dex even at solar metallicity. 
Simulations
do not predict observed drop in [$\alpha$/Fe] which is the signature of SNIa to the thick disk
chemical enrichment (see Figures 1 $\&$ 2). 

Radically different from the merger/accretion scenarios, recent studies focused 
on creating the
thick disk out of the Galactic disk without accretion from outside. Radial migration of stars
from inner to outer radii and the scattering of stars by spiral structure and the molecular
clouds could explain many of the observed morphological, kinematical and chemical properties
of the thick disk (e.g; Haywood 2008, Schonrich $\&$ Binney 2009). 
Though, this is an
interesting exercise, there are many assumptions like rate of infall and its distribution, and
scattering time scales that need to be constrained further. Models based on radial
mixing may find it difficult to explain the evidence of counter rotating 
thick disk stars that are seen in some of the spiral galaxies.

\section{Conclusions}
Significant progress has been made, over the last two decades, towards establishing various
observational properties. However, we still lack information
on the metal-poor side of the thick disk. This has implications
to favour a particular merging model. 
The issue of understanding the MWTD is related to accurate astrometry for a large number
of stars at relatively farther distances. Hopefully, this would
be resolved with the proposed GAIA space mission data expected to be available
in 2015/16.
Systematic studies of 
thick disk in large
number of spiral galaxies with varying mass would help 
to constrain the mass of the merging satellites in the simulations
of a particular galaxy with a particular thick disk to thin disk mass ratio.
As far as decoding the disk
components in other galaxies, we require more powerful tools like next generation
large aperture telescopes which are 
at the horizon. As shown in Figure~2, thick disk structure is much more complex
and observations point to multiple process for this structure: mergers, accretion of stars from the debris,
heating of the thin disk through scattering, and radial mixing.  Hopefully, in the near future, an unified
model will be developed that would match all the well defined observed parameters.

\begin{acknowledgments}
I thank SOC for inviting and giving me this opportunity to review the thick disk
and LOC for their superb arrangements at the
meeting. I thank INSA, IAU and IIA for providing the funds to attend the meeting.   
Also,
I thank T. Sivarani and G. Pandey for their useful comments.
\end{acknowledgments}

\end{document}